# Phase retrieval of programmable photonic integrated circuits based on an on-chip fractional-delay reference path

Xingyuan Xu,[1, 2, #] Guanghui Ren,[3, #] Aditya Dubey,[2] Tim Feleppa,[1] Xumeng Liu,[1] Andreas Boes,[3, 4, 5, *] Arnan Mitchell,[3] Arthur J. Lowery[1, *]

[1]Electro-Photonics Laboratory, Department of Electrical and Computer Systems Engineering, Monash University, Clayton, 3800 VIC, Australia

[2]State Key Laboratory of Information Photonics and Optical Communications, Beijing University of Posts and Telecommunications, Beijing 100876, China

[3]Integrated Photonics and Applications Centre, School of Engineering, RMIT University, Melbourne, 3000 VIC, Australia

[4]Institute for Photonics and Advanced Sensing (IPAS), University of Adelaide, Adelaide, 5005 SA, Australia

[5]School of Electrical and Electronic Engineering, University of Adelaide, Adelaide, 5005 SA, Australia

[#]These authors contribute equally to this work.

*Corresponding author: andy.boes@adelaide.edu.au, arthur.lowery@monash.edu

## Abstract

Programmable photonic integrated circuits (PICs), offering diverse signal processing functions within a single chip, are promising solutions for applications ranging from optical communications to artificial intelligence. While the scale and complexity of programmable PICs is increasing, the characterization, and thus calibration, of them becomes increasingly challenging. Here we demonstrate a phase retrieval method for programmable PICs using an on-chip fractional-delay reference path. The impulse response of the chip can be uniquely and precisely identified from only the insertion loss using a standard complex Fourier transform. We demonstrate our approach experimentally with a 4-tap finite-impulse-response chip. The results match well with expectations and verifies our approach as effective for individually determining the taps' weights without the need for additional ports and photodiodes.

**Introduction**

Programmable photonic integrated circuits (PICs), which are assemblies of reconfigurable integrated optical elements, support the high bandwidths needed for signal processing [1-8]. Being empowered by the reconfigurability of optical building blocks, they enable diverse functions within a single chip, such as routing, switching and equalizing for optical communications, and spectrum slicing and filtering for signal processing; thus, they are promising for a wide range of applications, including optical communications and signal processing [9-16], quantum computing [17], and artificial intelligence [18].

The specific functionality of a linear PIC is dictated by its transfer function (i.e., frequency response) or equivalently its temporal impulse response. Owing to the difficulties in maintaining sub-wavelength accuracies during fabrication [19], adaptive post-production tuning is commonly necessary to deliver a desired frequency response. Furthermore, calibrated tuning can support tunable filtering and routing. While the scale and complexity of programmable PICs is increasing, supported by maturing fabrication processes, the difficulties in accurately characterizing and thus controlling programmable PICs is becoming a severe limitation to their practical application. This is mainly due to the challenge in accurately calibrating the phase response of an optical chip, especially when the optical chip is included into an external interferometer such as a two-port optical analyzer, and connected via flexible patch leads.

Recently, we proposed an approach to accurately calibrate the phase response of an optical chip that builds an on-chip interferometer by adding an on-chip reference path, around the desired signal processing core [20]. The reference path could be an additional arm of a finite-impulse-response (FIR) filter, for example, but with the shortest delay. The added reference path enables the Kramers-Kronig relationship [21-25] to be used to recovery the phase information of the chip from its power response; this response is easy to measure using a tunable laser source (which are ubiquitous in wavelength-division multiplexing communications systems) and a photodiode, both of which could be on-chip or off-chip, in which case, the technique is immune to phase variations in the patch leads. Our demonstrated chip calibration technique provides a stable and accurate approach for the characterization of the chip's full frequency response, including optical phase, under various tuning conditions. However, the Kramers-Kronig phase retrieval requires certain conditions for it to be valid, often stated as the Minimum Phase condition. This condition can be guaranteed by using a very strong signal in the reference path, or more precisely, a signal stronger in power than the signal in the signal processing core. Although the Minimum Phase condition can be straight-forwardly implemented for generic applications, the relatively weak optical power allocation to the signal processing core could lead to limitations onto the signal-to-noise ratio for calibration.

In this paper, we demonstrate a phase retrieval method for programmable PICs based on adding an on-chip fractional-delay reference path. We show that if the reference path has half the delay of the delay increments in the signal processing core, the phase and amplitude weights of every delay path in the chip can be uniquely and precisely identified from only the power response (i.e.

containing no phase information) using a standard complex Fourier transform. In contrast to our previous work that used Kramers-Kronig relations, this method removes the optical power restrictions brought about by the Minimum Phase condition, thus enabling a potentially higher signal-to-noise ratio in the signal path for calibration; and eliminates the need of Hilbert transforms and nonlinear functions to reduce the computational complexities of the calibration system. We performed detailed theoretical simulations of three generic signal processing functions, including a 32-tap FIR filter, a micro-ring resonator (MRR), and cascaded MRRs; and experimental demonstrations using a 4-tap FIR filter integrated on a chip. The results match well with expectations and verifies our approach as effective for the calibration of programmable PICs.

**Theory**

As shown in Figure 1, our custom PIC consists of a signal processing core coupled in parallel with a reference path that has the shortest delay on the chip. The signal processing core can be implemented in diverse forms such as finite impulse response (FIR) filters and infinite impulse response (IIR) filters [2], with a frequency response of $H_{spc}$ and a discrete impulse response of $h_{spc}$.

The operation of the phase extraction consists of two steps. In the first step, the insertion loss (i.e., the magnitude-squared amplitude response, or power response) within the calibrating free spectral ranges (FSRs) is measured with a wavelength-swept laser and an optical power meter, via the optical ports accessing both the signal processing core and the reference path. We shall show that this measurement is sufficient to recover the phase and amplitude weights of each delay path relative to the reference, and thus the frequency and impulse responses of the signal processing core. In the second step, after the chip is calibrated/reconfigured, the signal processing core can be used for subsequent applications via the signal input/output ports, in free-spectral ranges other than those used for calibration.

The power response $|H_{chip}|^2$ of the whole chip (i.e., the insertion loss spectrum that is straightforward to measure) can be regarded as the superposition of frequency-domain raised-sinusoidal responses originating from: (i) mutual interferences between pairs of the signal processing core's taps (i.e., the discrete complex impulse response); and (ii) the many interferences between the reference path and each of the signal processing core's taps.

By performing Fourier transform of $|H_{chip}|^2$, a temporal series $h_{rec}$ can be obtained, which contains: the impulse response information of the signal processing core corresponds to interferences (ii), and mutually interfered spurious terms corresponds to interferences (i). We show that by choosing the delay of the reference path carefully, we can separate the components (i) and (ii) and so isolate the impulse response of the signal processing core using an inverse Fourier transform of the intensity response spectrum.

Specifically, the temporal interval between the reference path and the shortest delay of the signal processing core is set as $T/2$, where $T$ is the minimum temporal interval (i.e., delay step) of $h_{spc}$. This indicates that, the delays between the signal processing core's paths and the reference path are $(n+0.5)T$, where $n$ is an integer, and the signal processing core's discrete impulse response, or taps, locate at $(n+0.5)T$, as shown in Fig. 1, and all spurious signals fall at $nT$. We note that, two FSRs (each given by $1/T$) of the signal processing core are needed for phase retrieval, as the introduced reference path features a fractional delay of $T/2$ and thus making the free spectral range of the whole chip $2/T$.

Because of the choice of our reference delay, the mutual interferences (i) will fall at different delays to the wanted signals (ii), and so can be rejected. The recovered signal processing core's taps locate at 'fractional' delays of $(2n+1)T/2 = (n+0.5)T$, while the unwanted interference (a) locate at integral delays of $(2n+2)T/2 = (n+1)T$, with $n= 0, 1, 2, …$, as illustrated in Fig. 1. If the zero delay is also rejected, we will be left with the complex impulse response for the signal processing core, which can be transformed to the signal processing core's frequency responses including both the amplitude and phase information.

In this way, the signal processing core's complex tap coefficients, or impulse response, can be recovered from the amplitude response of the whole chip, enabling subsequent calibration and reconfiguration of the signal processing core towards desired functions. The integrated reference path offers a compact and stable solution for on-chip phase recovery in programmable PICs. In contrast to our previous work, this phase recovery approach further removes the requirement that for a large proportion of the input power to be used for the reference path, and the need for Kramers-Kronig phase recovery including its Hilbert transform and nonlinear functions, potentially enabling more robust and faster calibration of programmable PICs.

**Results**

To verify the reach of our approach, we first performed theoretical simulations of three typical on-chip signal processing building blocks, including a 32-tap FIR filter, a single MRR, and cascaded MRRs. As shown in Fig. 2, assisted by the reference path, the impulse response of the signal processing core can be accurately extracted from the power response via a single Fourier transform operation, verified by the close match between the recovered and ideal transfer function of the signal processing core, manifesting that our approach can serve as a universal phase recovery method for generic on-chip photonic devices. We note that the power spectrum is supplied directly from the real frequency-domain input of the Fourier transform, and the imaginary part is null.

We also fabricated a 4-tap FIR chip with a build-in reference on a silicon-nitride loaded thin-film lithium niobate on insulator ($Si_3N_4$/LNOI) platform [26, 27], as shown in Fig. 3. Electron-beam lithography and reactive ion etching were used to define the optical waveguides. Plasma enhanced chemical vapor deposition created a 2-µm thick silica cladding layer, and metal lift-off was used to define the micro thermal heaters to control the power splitting of the Mach-Zehnder interferometers and the phase of the delays.

We first verified the phase recovery process for a specific set of tap coefficients (in the case no electrical power is applied to the FIR chip). The insertion loss spectrum (i.e., squared amplitude response) of the whole chip $|H_{chip}(\omega)|^2$ was measured with an optical vector network analyzer (OVNA, Luna OVA 5000) (Fig. 3(c)), and used for the recovery of the complex impulse response (Fig. 3(e)) via the Fourier transform. The OVNA was also able to internally calculate the impulse response of the chip, including an impulse from the reference path. Fig. 3(e) demonstrates the close match between the recovered (blue dots) and measured (grey line) amplitudes of the impulse response, verifying our approach experimentally. The full complex frequency response of the FIR signal processing core can thus be calculated, and used to calibrate it for optical signal processing applications.

Further, we swept the electrical power onto Phase Shifter 1 (labelled in Fig. 3(a), corresponds to the phase of the FIR filter's first tap) from 0 to 0.2 W. As shown in Fig. 3(f, h), while the first tap's phase increases linearly with applied power, the amplitudes and phases of the remaining taps are almost constant, albeit with minor variations due to thermal cross-talk. These variations could be calibrated out with a suitable intelligent control algorithm. The initial phase of tap 1 was extracted as $0.272\pi$, and the electrical power needed to achieve $2\pi$ phase shift is 0.221 W.

**Conclusion**

In conclusion, we have demonstrated a phase retrieval method for programmable PICs. By employing a build-in reference path with fractional delay, the impulse response of the chip can be determined from the power response via a single Fourier transform. We performed detailed theoretical simulations of three generic signal processing function, including a 32-tap FIR filter, a micro-ring resonator (MRR), and cascaded MRRs; and experimental demonstrations using a 4-tap FIR filter integrated on a chip. The results verify our approach as effective for the calibration of programmable PICs, potentially enabling fast reconfiguring PICs towards applications ranging from optical computing to quantum applications.

**Acknowledgements:** We thank Mr. Andrew Linzner for the fabrication of metal chip holders. This work was supported by the Australian Research Council Discovery Projects Program (No. DP190101576, DP190102773).

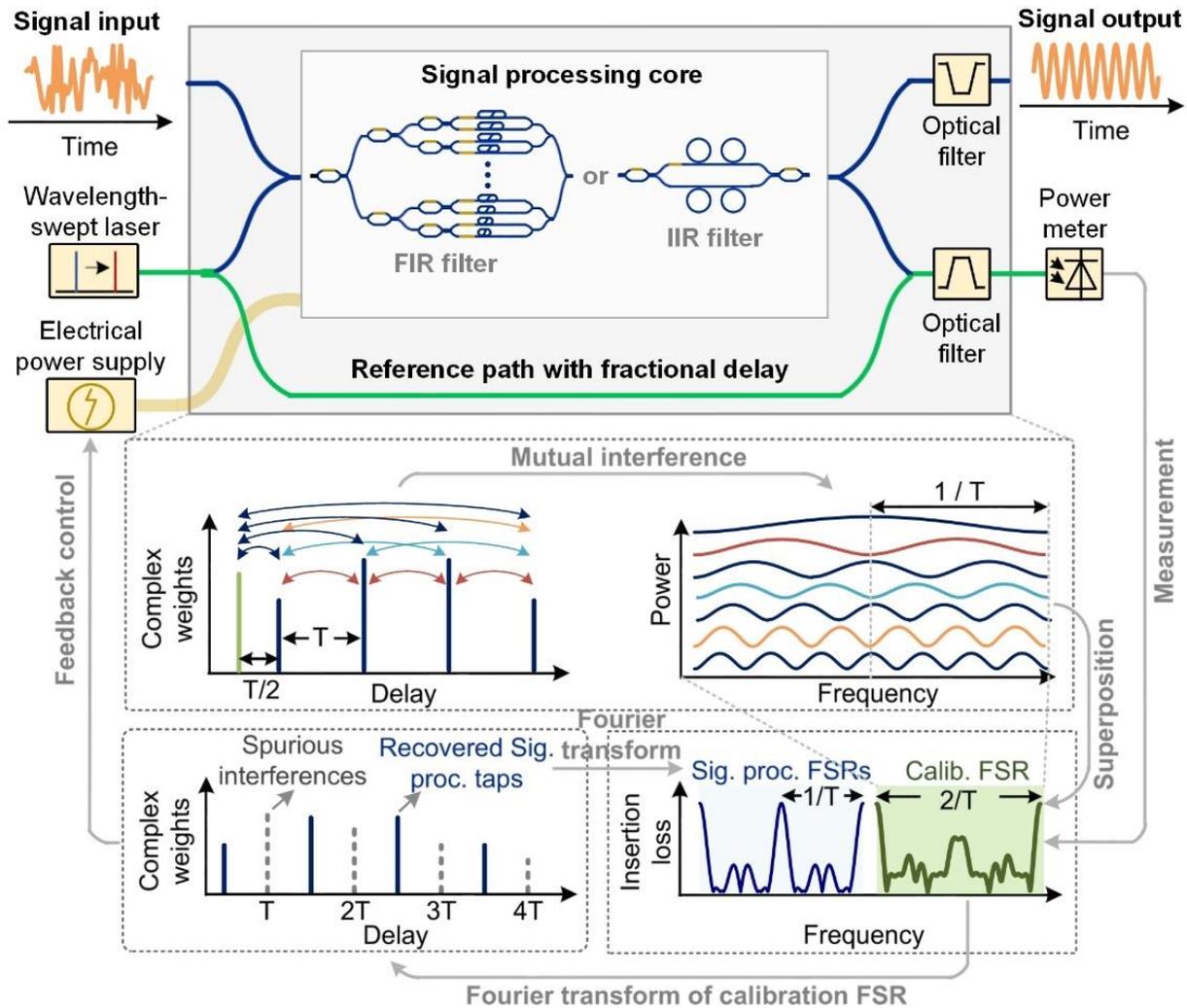

**Figure 1 | Conceptual diagram of the phase recovery approach based on a fractional-delay reference path.** The chip consists of a signal processing core and a reference path with fractional delay. The signal processing core can be implemented with diverse structures such as FIR or IIR filters. The chip features two pairs of optical input and output ports, including the top pair for signal processing (blue line) and the bottom pair for calibration (green line). A wavelength-swept laser and an optical power meter are used to measure the power response of the whole chip within two FSRs (denoted as 2/T), while the remaining FSRs are used for signal processing, guaranteed by the optical filters. The power response of the chip can be regarded as the superposition of frequency-domain raised-sinusoidal responses originating from mutual interferences all taps. The signal processing core's tap coefficients are recovered via a Fourier transform, then used for error calculation and feedback control of the electrical power applied onto the chip.

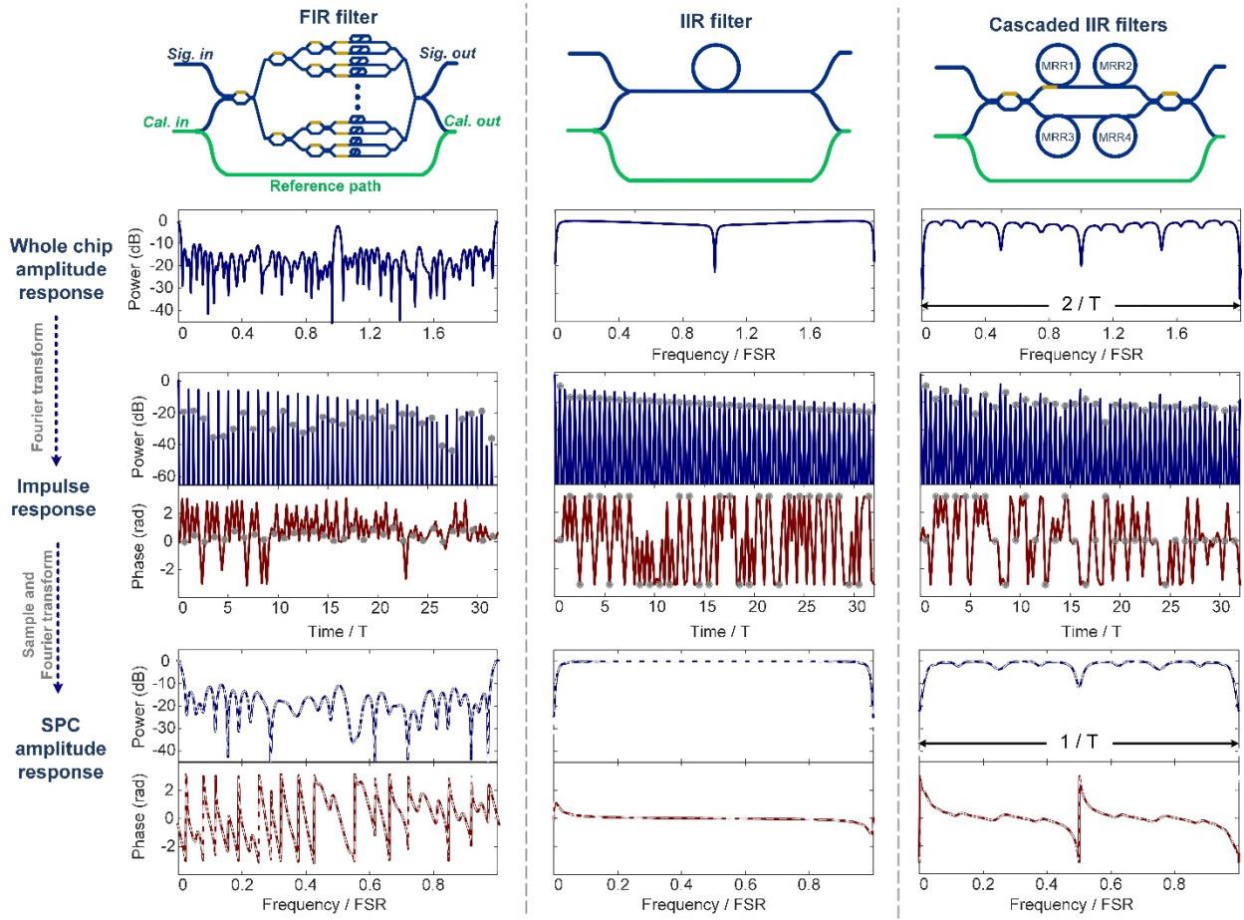

**Figure 2 | Simulated phase recovery results**, including three typical architectures for signal processing (top row): a 32-tap FIR filter with random tap coefficients, a micro-ring resonator, cascaded micro-ring resonators; the corresponding amplitude response of the whole chip (second row); the recovered impulse responses after Fourier transform (solid line, third row), where the grey dots and circles show the ideal and recovered values, respectively; the corresponding transfer function of the signal processing core calculated from the recovered impulse responses (solid line, bottom row), where the white dashed lines, indicating the ideal values calculated from the ideal tap coefficients, overlap with the recovered results.

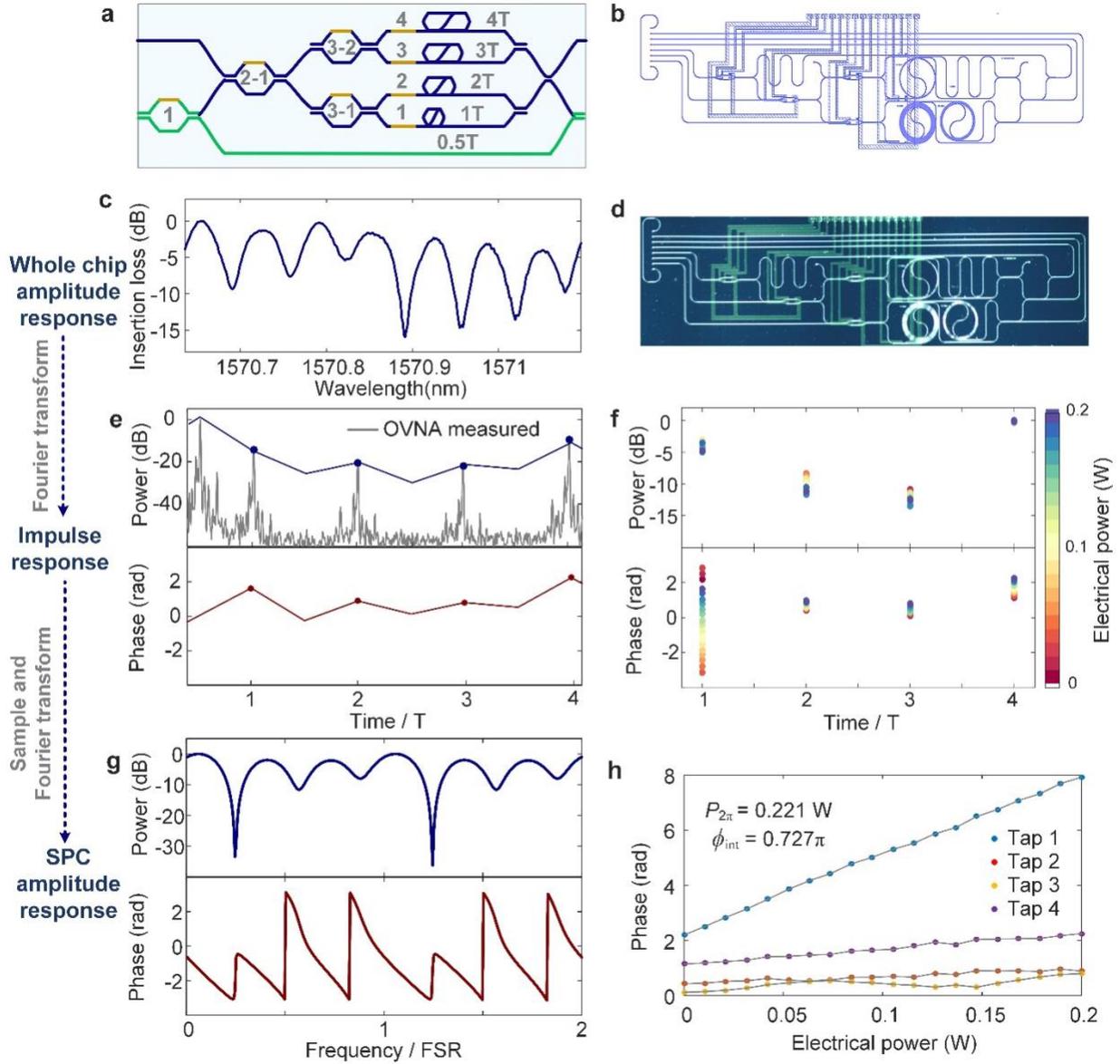

**Figure 3 | Experimental phase recovery results of a 4-tap FIR filter.** (a) shows the chip's architecture, where $T$ = 35.8 ps, corresponding to an FSR of $1/T$ = 28 GHz for the FIR filter; (b) shows the chip's layout; (d) photomicrograph. The phase recovery process for a specific set of tap coefficients is shown in (c, e, g): (c) shows the measured insertion loss spectrum after normalisation (i.e., power or squared amplitude response) of the whole chip – note the delay increment and the delay between the reference and first path's tap; (e) shows the recovered impulse response (blue/red lines and dots) and the measured impulse response' amplitudes using a commercial equipment (grey line); (g) shows the transfer function of the FIR signal processing core calculated from the recovered impulse response. The electrical power of the first tap was swept, with the recovered impulse responses shown in (f, h): (f) the powers and phases of the 4 taps as a function of the electrical power applied onto Phase Shifter 1; (g) the phases of the 4 taps as a function of the electrical power.